\documentclass{article}

\PassOptionsToPackage{numbers}{natbib}

\usepackage[preprint]{neurips_2023}

\usepackage[utf8]{inputenc} 
\usepackage[T1]{fontenc}    
\usepackage{natbib}
\usepackage{hyperref}       
\usepackage{url}            
\usepackage{booktabs}       
\usepackage{amsfonts}       
\usepackage{nicefrac}       
\usepackage{microtype}      
\usepackage{xcolor}         
\usepackage{bm}
\usepackage{amsmath}
\usepackage{amssymb}
\usepackage{graphicx}
\hypersetup{    
  colorlinks=true,
  linkcolor=red,  
  citecolor=blue, 
  urlcolor=blue    
}

\definecolor{mygreen}{RGB}{20,138,6}
\definecolor{myblue}{RGB}{52,180,230}
\definecolor{mypurple}{RGB}{160,60,160}

\title{A deep learning framework for jointly extracting spectra and source-count distributions in astronomy}

\author{%
  Florian Wolf$^1$ $\qquad$ Florian List$^1$ $\qquad$ Nicholas L. Rodd$^2$ $\qquad$ Oliver Hahn$^1$ \\
  $^1$University of Vienna, Department of Astrophysics \& Department of Mathematics \\  
  \texttt{\{florian.wolf, florian.list, oliver.hahn\}@univie.ac.at} \\
  $^2$CERN, Theoretical Physics Department \\
  \texttt{nick.rodd@cern.ch}  
}

\begin{document}

\maketitle

\begin{abstract}
Astronomical observations typically provide three-dimensional maps, encoding the distribution of the observed flux in (1) the two angles of the celestial sphere and (2) energy/frequency.
An important task regarding such maps is to statistically characterize populations of point sources too dim to be individually detected.
As the properties of a single dim source will be poorly constrained, instead one commonly studies the population as a whole, inferring a source-count distribution (SCD) that describes the number density of sources as a function of their brightness.
Statistical and machine learning methods for recovering SCDs exist; however, they typically entirely neglect spectral information associated with the energy distribution of the flux.
We present a deep learning framework able to jointly reconstruct the spectra of different emission components and the SCD of point-source populations.
In a proof-of-concept example, we show that our method accurately extracts even complex-shaped spectra and SCDs from simulated maps.
\end{abstract}

\section{Introduction}
\label{sec:intro}

In many senses, the most fundamental problem in astronomy is to provide a complete catalog of point sources (PSs) within a map of the sky.
Individual sources are traditionally characterized down to a detection threshold intrinsic to the instrument which collected the map.
For sources below this boundary, characterizing even basic properties such as their location becomes less meaningful; however, properties of the entire population of dim sources can be reliably inferred at a \textit{statistical} level, through, for instance, a source-count distribution (SCD), which reports the distribution of sources as a function of brightness.

Likelihood-based statistical methods have been applied to extract SCDs at various wavelengths---radio, optical, X-ray, gamma-ray, and even in high-energy neutrinos datasets, see e.g.~
\cite{IceCube:2019xiu, 1992ApJ...396..460B, Collin2022,Lee2016, Malyshev:2011zi, Mishra-Sharma2017, Miyaji:2001dp, 1957PCPS...53..764S, Zechlin2016, Zechlin2016a}. %
In principle; these methods are optimal for perfectly modeled data (apart from inter-pixel correlations introduced by the instrument point-spread function (PSF) \cite{Collin2022}); however, data mismodeling -- present to a certain degree in every analysis -- biases the results (e.g.\ \cite{Buschmann2020, Chang2019, Leane2019a, Leane2020, Leane2020a}.) 
Recently, deep learning methods have been developed, with a particular focus on gamma-ray astronomy and the data collected by the \emph{Fermi} satellite \cite{Atwood2009} (e.g.\ \cite{butter2023searching, Caron2022, Caron2018, List2020b}), several of which can determine SCDs from the {\it Fermi} photon-count maps \cite{amerio2023extracting, List2021, Mishra-Sharma2022}. 
Whilst these deep learning methods have their own systematics (see e.g. \cite{List2021} for an exploration), they also hold the potential of being able to extend beyond what was possible with likelihood-based approaches.

To be explicit, an important limitation of the likelihood approach is that it is by construction blind to any energy or frequency information, studying only the distribution of flux on the celestial sphere.
In principle, one can write the likelihood with the energy dimension included, but evaluating this proves to be a significant combinatorial problem that remains unresolved \cite{Collin2022}.
Deep learning methods, by contrast, should not be limited in this manner: the input data can be provided binned in energy, allowing for basic quantities like the spectrum (energy distribution) of sub-threshold PS populations to be inferred, without the need for an analytic expression of the likelihood.
Despite this promise, existing deep learning approaches have generally ignored energy information.

In this work, we present a first demonstration of a deep learning framework that enables simultaneous inference of spectra and SCDs from astronomical data, thereby exploiting the full information collected by the telescope. We choose to focus on gamma-ray astronomy given the wide class of techniques developed at that wavelength, but we emphasize that our approach could be readily extended to other energies or messengers.

\section{Methods}
\label{sec:methods}

\subsection{Modeling}
\label{subsec:modeling}
We consider photon-count maps (e.g. of the entire sky) $\bm x = (x_n^e)_{n=1,\ldots,N_{\mathrm{pix}}}^{e=1,\ldots,E} \in \mathbb{N}_{0}^{N_{\mathrm{pix}} \times E}$, which contain the counts detected in $E$ energy bands in each of the $N_{\mathrm{pix}}$ pixels.\footnote{Herein, we ignore the sky exposure and use the notions of ``counts'' and ``flux'' interchangeably.} We model these counts as originating from $T$ emission components, each of which has an individual spatial morphology and spectrum, i.e.\ $\bm x = \sum_{t = 1}^T \bm x^t$, where $\bm x^t \in \mathbb{N}_{0}^{N_{\mathrm{pix}} \times E}$ are the counts from component $t$.

We distinguish between two types of emission components:
\begin{enumerate}
    \item \textbf{Poissonian} components (e.g.\ different foregrounds), for which we assume the spatial morphology to be known so that the counts associated with such a component $t$ are simply described by a Poisson draw $\bm x^{t,e} \sim \operatorname{Pois}(A^{t} \, \xi^{t, e} \, \bar{\bm x}^{t,e})$. Here, $\bar{\bm x}^{t,e} \in \mathbb{R}_{\geq 0}^{N_{\mathrm{pix}} \times E}$ models the (known and normalized) spatial morphology of the counts in each pixel and energy bin, $A^{t}$ is an (unknown) normalization coefficient that characterizes the total brightness of the component, and $\xi^{t, e} \in [0, 1]$ with $\sum_{e=1}^E \xi^{t, e} = 1$ encodes its (unknown) spectrum. 
    \item \textbf{Point-source} components, for which the situation is somewhat more complicated: here, the underlying spatial distribution $p_{\mathrm{PS}}^t$ of the PSs belonging to component $t$ is assumed to be known, but not the individual location of each source. In this proof-of-concept paper, we take the PSs to be uniformly distributed across the sky for simplicity. So for the (unknown) longitude $l_s^t$ and latitude $b_s^t$ of each source $s^t$, one has $(l_s^t, b_s^t) \sim p_{\mathrm{PS}}^t$ (which is of course the same for all energy bands $e$). The total number of sources is modeled as a Poissonian random variable $N_{\mathrm{PS}}^t \sim \operatorname{Pois}(\bar{N}_{\mathrm{PS}}^t)$, where the expected number of sources $\bar{N}_{\mathrm{PS}}^t$ associated with component $t$ is unknown. Each source $s^t$ emits a certain number of counts $c_s^{t, e}$ in each energy band $e$ according to a Poisson process $c_s^{t, e} \sim \operatorname{Pois}(\xi^{t, e} \, \mu_s^{t})$, where $\mu_s^{t} \sim \mathrm{SCD}$ reflects the brightness of that source (integrated over all energy bands), and $\xi^{t, e} \in [0, 1]$ with $\sum_{e=1}^E \xi^{t, e} = 1$ characterizes the spectrum (which we take to be the same for all sources belonging to component $t$). The counts $c_s^{t, e}$ from each source $s^t$ are localized around the pixel associated with the sky coordinates $(l_s^t, b_s^t)$, but smeared within a vicinity of this pixel due to the (energy-dependent) instrument PSF. The distribution of the values $\mu_s^t$ (i.e., the SCD of component $t$) and the relative spectrum $\xi^{t, e}$ are both unknown.
\end{enumerate} 
Our framework simultaneously targets two objectives: (1) the inference of the \textbf{spectra} as described by the spectral shape $\xi^{t, e}$ and an appropriate normalization for each component (both Poissonian and point-like), and (2) the inference of the \textbf{SCDs} of all PS components, i.e.\, the distribution of the source brightnesses $\mu_s^{t}$.

\subsection{Extracting spectra}
\label{subsec:spectra}
For recovering spectra from the data, we extend the Gaussian maximum likelihood estimation proposed by Ref.~\cite{List2020b} for the inference of count fractions to multiple spectral bands. For simplicity, we neglect the Poisson scatter when estimating spectra and use the actually {\it realized} counts, rather than the underlying spectral shapes $\xi^{t, e}$ for the labels, as the total number of counts for each component is typically large. Inferring the spectra in terms of counts per energy bin simplifies the task, as this leads to a unified approach for all components (Poissonian and point-like).

Since the total number of detected counts in each energy band summed over all components is known from observation, extracting the spectra is tantamount to estimating the \textit{fraction} of counts contributed by each component, separately for each energy bin.

Given a map $\bm x$, we therefore task a neural network (NN) with trainable parameters $\bm \theta$ to provide estimates $\tilde{\bm y}_{\bm \theta} = (\tilde{y}^{t,e}_{\bm \theta}) \in [0, 1]^{T \times E}$ for the relative counts of all components $t$ in each energy bin $e$,
\begin{equation}
    \tilde{y}^{t, e}_{\bm \theta} \approx y^{t,e} = \left(\sum_{n=1}^{N_{\mathrm{pix}}} x^{t, e}_n\right) \Bigg{/} \left(\sum_{n=1}^{N_{\mathrm{pix}}} x^{e}_n\right)\!.
    \label{eq:count_fraction}
\end{equation}
We train the NN to estimate posteriors by minimizing the negative Gaussian log-likelihood
\begin{equation}
    \mathcal{L}_{\mathrm{spectrum}}(\tilde{\bm y}_{\bm \theta}(\bm x), \bm y) = \sum_{t = 1}^T \sum_{e = 1}^{E} \left(\frac{1}{2 (\tilde{\sigma}^{t,e}_{\bm \theta}(\bm x))^2} \left(\tilde{y}^{t, e}_{\bm \theta}(\bm x) - y^{t, e}\right)^2 + \frac{\ln \left[(\tilde{\sigma}^{t,e}_{\bm \theta}(\bm x))^2 \right]}{2} \right)\!.
    \label{eq:loss_spectrum}
\end{equation}
The standard deviation $\tilde{\sigma}^{t,e}_{\bm \theta}(\bm x)$ associated with the aleatoric uncertainty of the prediction $\tilde{y}^{t, e}_{\bm \theta}(\bm x)$ is estimated by the NN itself, in such a way as to minimize Eq.~\eqref{eq:loss_spectrum}. This loss function assumes the count fractions of the different components in different energy bands to be uncorrelated (which is clearly an approximation, just as the assumption of Gaussianity); however, extending our method to a full covariance structure along the component and/or energy dimension is possible. Another interesting avenue -- which we will investigate in future work -- is weighting the residual in each energy bin by the total number of counts in that bin, in order to penalize the NN in proportion to its prediction error in terms of total rather than relative counts. Once the NN has been trained, the spectrum of each component $t$ in terms of total counts per energy bin $e$ is readily obtained from $\tilde{\bm y}_{\bm \theta}(\bm x)$ by solving for $\sum_{n=1}^{N_{\mathrm{pix}}} x_n^{t,e}$ in Eq.~\eqref{eq:count_fraction}.

\subsection{Extracting source-count distributions}
\label{subsec:scds}
For extracting the SCDs of the PS components, we adopt the non-parametric approach proposed by Ref.~\cite{List2021} and discretize the SCD of each PS component as a histogram $\bm u = (u^m) \in [0, 1]^M$, where $M$ is the total number of (logarithmic) count bins (which is a hyperparameter). 
As mentioned above, we assume the SCDs to be energy-independent and infer a single SCD for each PS component.  
Therefore, the only difference in comparison with the framework in that reference is that the input maps $\bm x \in \mathbb{N}^{N_{\mathrm{pix}} \times E}$ on which the NN is trained are energy-dependent, but the final recovered SCDs remain energy-independent, so we will only summarize the key points here. 

For computing the SCD value $u^m$ in bin $m$, we sum up the expected counts contributed by all those PSs whose expected number of counts falls within the $m$-th count bin; then, we normalize the histogram to sum up to unity, i.e.\ $\sum_{m=1}^M u^m = 1$. This is because the normalization in terms of total counts associated with each component can be obtained from the estimated spectra as $\sum_{e=1}^E \sum_{n=1}^{N_\mathrm{pix}} x_n^{t, e}$, so estimating relative SCDs is sufficient.

For extracting the SCDs, we use a second NN with parameters $\bm \vartheta$, trained with a loss function rooted in quantile regression \cite{Koenker1978}. Specifically, we apply the pinball loss \cite{fox1964admissibility, Koenker1978} to the {\it cumulative} histogram with $U_m := \sum_{j=1}^m u_j$ in each bin, i.e.\ \cite{list2021earth}
\begin{equation}
    \mathcal{L}_{\mathrm{SCD}}^\tau(\tilde{\bm u}_{\bm \vartheta}(\bm x; \tau), \bm u) = \frac{1}{M} \sum_{m=1}^M\left[\left(\tilde{U}^m_{\bm \vartheta}(\bm x; \tau) - U^m\right)\left(\tau-\mathbb{I}\left[\tilde{U}^m_{\bm \vartheta}(\bm x; \tau) < U^m\right]\right)\right]\!,
\end{equation}
where $\tilde{\bm u}_{\bm \vartheta}(\bm x; \tau)$ is the SCD histogram estimated by the NN at quantile level $\tau$ with cumulative counterpart $\tilde{\bm U}_{\bm \vartheta}(\bm x; \tau)$, and $\mathbb{I}[\cdot]$ is the indicator function. The queried quantile level $\tau \in (0, 1)$, which is fed as an additional input to the neural network, is drawn randomly as $\tau \sim U(0, 1)$ during training and can be specified at inference time. As this loss function is minimized by the $\tau$-th quantile level, it enables us to estimate the {\it distribution} of possible SCDs, i.e.\ a median estimate ($\tau = 0.5$, in which case the loss function reduces to the $\ell^1$-loss applied to the cumulative histograms) plus uncertainties.

\subsection{Deep learning model}
We use two separate NNs stacked on top of each other, one for extracting the spectra, and another one for inferring the SCDs. The reason for sequentially training two NNs is that, in principle, the spectral information extracted by the first NN could be exploited by the second NN tasked with the SCD inference; we will investigate this in future work. To accommodate large regions of interest (ROI) for which the sky curvature becomes relevant, we build our NNs onto the \texttt{DeepSphere} architecture (\cite{Defferrard2020, Perraudin2019a}, see \cite{List2021, List2020b, Mishra-Sharma2022} for gamma-ray applications), which models the sky as a graph based on the \texttt{Healpix} tessellation of the sphere \cite{2005ApJ...622..759G} and provides graph counterparts of convolution and pooling.

\section{Results}
\label{sec:results}
To demonstrate the effectiveness of our method, we consider photon counts in a $25^\circ$ ROI. The photons originate from an isotropic PS component and a smooth (Poissonian) background component (i.e., $T = 2$). We model the smooth background with the pion decay \& bremsstrahlung component of the \texttt{Model O} template presented in Ref.~\cite{Buschmann2020} (building on Refs.~\cite{2018NatAs...2..387M, 2019JCAP...09..042M}). Although we avoid some of the technicalities present in a realistic analysis (such as the non-uniform sky exposure by gamma-ray telescopes etc.), this experiment comprises the major difficulties one encounters in practice, namely the presence of multiple overlapping emission components (with a background whose morphology varies as a function of energy) and a \emph{Fermi}-like PSF that deteriorates with decreasing energy, described by a combination of King functions.\footnote{\url{https://fermi.gsfc.nasa.gov/ssc/data/analysis/documentation/Cicerone/Cicerone_LAT_IRFs/IRF_PSF.html}} 

We model the SCD and the spectra of the isotropic PSs as a mixture of two truncated skew-normal distributions whose location, scale, and skewness parameters are drawn from wide distributions. The motivation for considering a mixture distribution is to highlight the ability of our framework to extract complex spectra and SCDs and, from a physical point of view, to account for the possibility that two distinct (in terms of spectrum and SCD) PS populations with a similar spatial morphology are hidden in the data. For the Poissonian background, we draw the spectrum from a single truncated skew-normal distribution.
We discretize the SCDs into $M = 22$ count bins and use $E = 10$ energy bins for the spectra.
Our training dataset consists of 940,000 simulated maps at \texttt{Healpix} resolution parameter $N_{\mathrm{side}} = 256$, and we train our NNs to infer the spectra of both components and the SCD of the PS population. We remark that our SCD estimation framework directly carries over to multiple PS populations with different spatial morphologies (as in Ref.~\cite[Fig.~8]{List2021}).

\begin{figure}
    \centering
    \includegraphics[width=\textwidth]{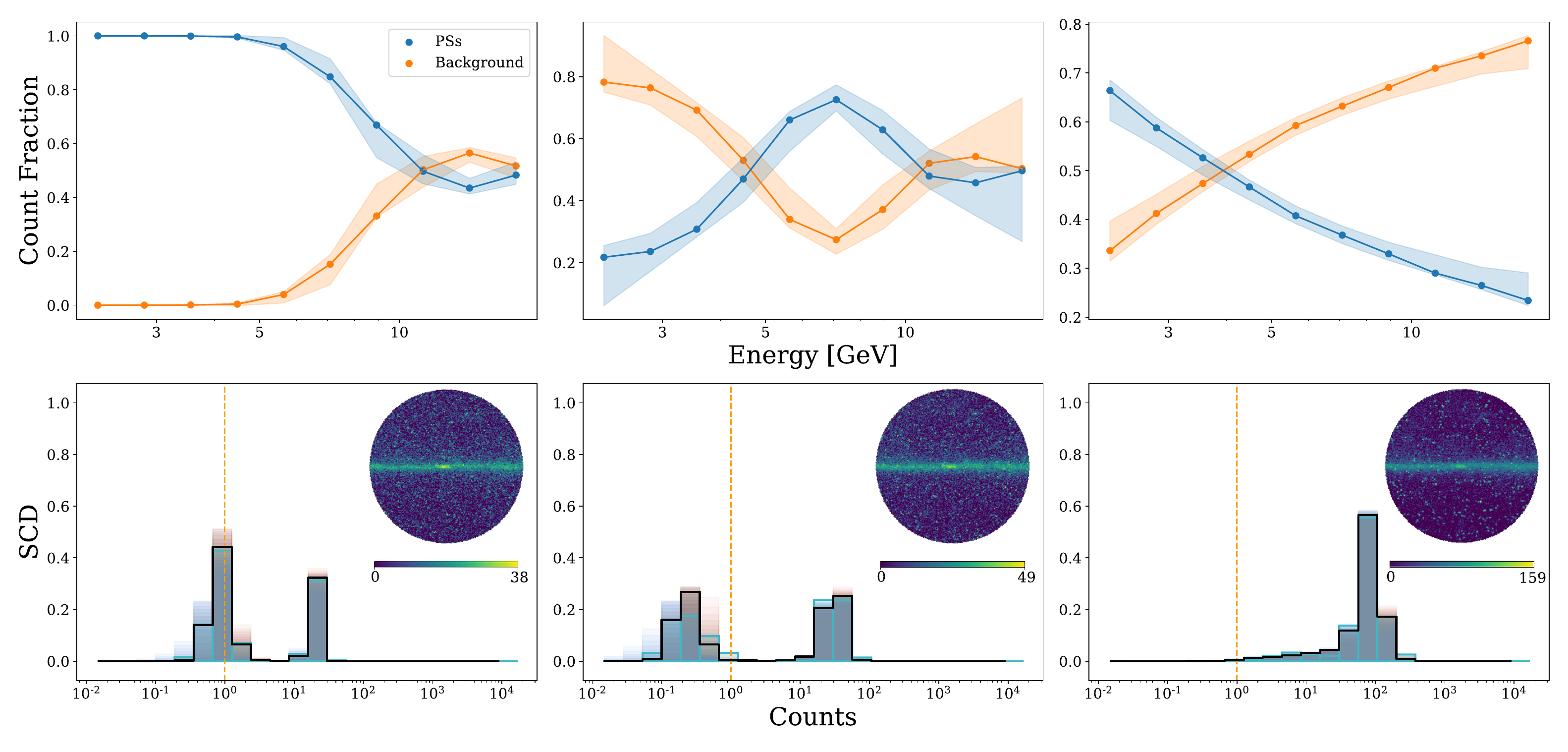}
    \caption{NN predictions for three selected test maps. \textit{Top:} Count fractions for the isotropic PS and the diffuse background component as a function of photon energy. The solid lines represent the true values, whereas the shaded areas cover the 1$\sigma$  NN predictions. 
    \textit{Bottom:} True vs. reconstructed SCD of the isotropic PS component. Here, the black line is the median prediction (bin-wise in terms of the associated cumulative histogram, see Sec.~\ref{subsec:scds}), while the faint colored regions indicate the quantiles from 5\% to 95\%. The blue line shows the true SCD. The orange vertical line corresponds to a single expected photon count per source. In the top right corner, we show the maps (summed over all energy bins) to which the spectra and SCDs belong. The color map is scaled logarithmically. 
    The SCDs are modeled as a mixture of two skew-normal distributions, and their bimodality is clearly visible.}
    \label{fig:results}
\end{figure}

Figure~\ref{fig:results} shows the spectra in terms of relative counts, i.e.\ $\bm y$ in Eq.~\eqref{eq:count_fraction}, and SCDs for three maps from the test dataset. We selected these maps in order to showcase some diverse and instructive configurations for the spectra and SCDs, not based on the performance of our NNs. We will present a thorough quantitative analysis and an application of our method to \emph{Fermi} data elsewhere. 

The NN accurately recovers the complex shapes of the count fractions which arise from the interplay between the different spectral shapes. The true spectra approximately fall within the 1$\sigma$ region for the maps shown in the figure, and the performance on other maps is similar. 

As for the SCDs, the NN is able to obtain accurate estimates for populations above the one photon line (dashed orange). Below, predictions become less precise: in view of PS emission being degenerate with Poisson emission in the limit of $\ll 1$ expected counts per source, exactly pinpointing the location of the SCD in this regime is no longer possible from a statistical point of view, see e.g. Ref.~\cite[Fig.~9]{List2021}. The first and second panel show maps for which the bimodality of the SCD clearly indicates the presence of a faint and a brighter PS subcomponent (described by the mixture of two skew-normal distributions, see above). Our NN produces larger uncertainties for the faint subcomponent (which our NN expresses in terms of the {\it cumulative} SCDs that we converted to density histograms, indicated by the blue-to-red colored regions), just as expected.

\section{Conclusions}
\label{sec:conclusions}
We introduced a machine-learning framework for the simultaneous extraction of SCDs and spectra from astronomical data, where we make use of the photon energy information.  Including spectral information in the analysis enables more accurate modeling on one hand (as all analysis components such as spatial templates for the different emission components, the instrument PSF, masks for known PSs etc.\ can be made dependent on the energy), and more informative output on the other hand, as it allows the model to infer the spectra in addition to the SCDs of the PS components. Also, physically motivated tight priors on the spectra should enable the NNs to obtain tighter constraints as they can learn the correlations between different energy bins; we will carry out a systematic exploration study in this direction in future work. Further, the method could be improved by incorporating the covariance between components or energies in the training.

There are many interesting applications for our method in practice, in particular for {\it Fermi} data. For instance, there has been an ongoing discussion in the field regarding the composition of the high-latitude sky in the GeV range, specifically in terms of the contributions of star-forming galaxies \cite{roth2021diffuse} and blazars (e.g.\ \cite{ajello2015origin, Manconi2020}). Also, the origin of the Galactic Center Excess \cite{goodenough2009, HOOPER2011}, for which the most popular contenders are a faint population of millisecond pulsars (e.g.\ \cite{KAbazajian2011, Gautam_2022, gordonMacias2013}) or dark matter annihilation (e.g.\ \cite{Agrawal_2015, Calore_2015, gordonMacias2013}), is still unknown. Jointly exploiting spectral properties and SCDs is crucial for further progress on both fronts.

Our method is also ideally suited for applications in X-ray (or neutrino) astronomy, where one often has to deal with complex ROIs and non-isotropic PSFs (e.g.\ \cite{harrison2013nuclear}), which poses difficulties for likelihood methods, but is straightforward for simulation-based approaches. There are also clear targets for the future. To begin with, in isolated cases we have observed a marked degradation in the performance of the spectrum reconstruction, and improving performance in those cases will be important before turning to real data. Further, the leading systematic uncertainty for template-based methods is the use of imperfect background models, or equivalently background models that are missing an emission component. Quantifying the performance of our approach in the presence of such imperfections will allow us to determine the systematic uncertainties associated with inferences drawn from actual astrophysical maps.

{
{\footnotesize The computational results presented have been achieved using the Vienna Scientific Cluster (VSC).}

\small
\bibliographystyle{abbrv}

\bibliography{neurips_2023}
}

\appendix

\section{Details of the neural network architecture and training}

We used an Adam optimizer \cite{kingma2014adam} with a learning rate of $5 \times 10^{-4}$ decaying exponentially at a rate of 0.99985 per iteration. The training dataset consisted of 940,000 maps, and we trained with a batch size of 256 for 80,000 iterations on a single NVIDIA A100 GPUs on the ``Vienna Scientific Cluster 5 (VSC5)''. The NN architecture can be seen in Tables \ref{tab:NN_ff} (spectra) \& \ref{tab:NN_SCD} (SCDs). In total, both NN combined have 6,252,480 parameters from which 6,246,464 are trainable and 6,016 are non-trainable.

\begin{table}[h]
    \centering
    \begin{tabular}{c c c c c}
    \hline
    \textbf{Layer} & \textbf{Operations} & \textbf{Output shape} & \textbf{Output $n_{side}$} & 
    \textbf{Parameters} \\
    \hline

      I & Input map (normalised)  &  32,892$\times$10 & 256  &  -  \\
      II & ConvBlock & 8,396$\times$32 & 128 & 1,696 \\
      III & ConvBlock & 2,212$\times$64 & 64 & 10,432 \\
      
    IV & ConvBlock & 606 $\times$ 128 & 32 & 41,344 \\
    V & ConvBlock & 164 $\times$ 256 & 16 & 164,608 \\
    VI & ConvBlock & 50 $\times$ 256 & 8 & 328,448 \\
    VII & ConvBlock & 14 $\times$ 256 & 4 & 328,448 \\
    VIII & ConvBlock & 4 $\times$ 256 & 2 & 328,448 \\
    IX & ConvBlock & 1 $\times$ 256 & 1 & 328,448 \\
    
    X & Append log$_{10} (S_{\mathrm{tot}})$ & 1 $\times$ 257 &  \\
    XI & ReLU $\circ$ FC & 1 $\times$ 2,048 & & 528,384 \\ 
    XII & ReLU $\circ$ FC & 1 $\times$ 512 & & 1,049,088 \\
    XIII & Reshape $\circ$ FC & 2 $\times$ 2 $\times$ 10 & & 20,480 \\
    XIV & Softmax (means only) & 2 $\times$ 2 $\times$ 10 & & - \\ 
    \hline
 \hline
 \\
    \end{tabular}
    \caption{Architecture of our NN for estimating the spectra. The ConvBlocks are made of graph convolutions, batch normalisation, ReLU activation and maximum pooling. In layer X we append the logarithmic total counts of the map. The final layer uses a softmax activation for the mean estimates and also returns the estimated uncertainty for each mean value. The output shape corresponds to 2 (means/variances) $\times$ $T (= 2)$ components $\times$ $E (= 10)$ energy bins.}
    \label{tab:NN_ff}
\end{table}

\begin{table}[h]
    \centering
    \begin{tabular}{c c c c c}
    \hline
    \textbf{Layer} & \textbf{Operations} & \textbf{Output shape} & \textbf{Output $n_{side}$} & 
    \textbf{Parameters} \\
    \hline
    
      I & Input map (normalised)  &  32,892$\times$10 & 256  &  -  \\
      II & ConvBlock & 8,396$\times$32 & 128 & 1,696 \\
      III & ConvBlock & 2,212$\times$64 & 64 & 1,0432 \\
      
    IV & ConvBlock & 606 $\times$ 128 & 32 & 41,344 \\
    V & ConvBlock & 164 $\times$ 256 & 16 & 164,608 \\
    VI & ConvBlock & 50 $\times$ 256 & 8 & 328,448 \\
    VII & ConvBlock & 14 $\times$ 256 & 4 & 328,448 \\
    VIII & ConvBlock & 4 $\times$ 256 & 2 & 328,448 \\
    IX & ConvBlock & 1 $\times$ 256 & 1 & 328,448 \\
    
    X & Append log$_{10} (S_{\mathrm{tot}})$ & 1 $\times$ 257 &  \\
    XI & Append $\tau$ & 1 $\times$ 258 &  \\
    XII & ReLU $\circ$ FC & 1 $\times$ 2,048 & & 530,432 \\ 
    XIII & ReLU $\circ$ FC & 1 $\times$ 512 & & 1,049,088 \\
    XIV & FC & 1 $\times$ 22 & & 11,264 \\
    XV &  Normalized softplus & 1 $\times$ 22 & & - \\ 
    \hline
 \hline
 \\
    \end{tabular}
    \caption{Architecture of our NN for estimating the SCDs. Note that in layer X we append the logarithmic total counts of the map, and in layer XI we append the quantile level $\tau$. The final layer uses a normalized softplus activation. The output shape corresponds to a $1$ (single PS component) $\times$ $M (= 22)$ count bins.}
    \label{tab:NN_SCD}

\end{table}


\end{document}